\documentclass[epj]{svjour}
\usepackage{amsmath,amssymb,graphics}
\begin{document}
\title{Skyrme-QRPA calculations for low-lying excitation modes 
in deformed neutron-rich nuclei}
\author{Kenichi Yoshida
}                     
%
%
\institute{Nishina Center for Accelerator-Based Science,
RIKEN, Wako, Saitama 351-0198, Japan}
%
\date{Received: 8 December 2008 / Revised version: 13 January 2009}
%
\abstract{
Low-frequency modes of excitation in deformed neutron-rich nuclei 
are studied by means of the quasiparticle random-phase approximation 
on the Skyrme-Hartree-Fock-Bogoliubov mean field. 
We investigate the microscopic structure of the soft $K^{\pi}=0^{+}$ modes 
systematically in neutron-rich Magnesium isotopes with $N=22, 24, 26$ 
and 28 close to the drip line, and it is found that 
the strong collectivity in $^{34}$Mg and $^{40}$Mg is acquired 
due to the coherent coupling between the $\beta$ vibration 
and the pairing vibration of neutrons. 
Microscopic structure of the $K^{\pi}=2^{+}$ modes 
changes gradually associated with the location of the Fermi level 
of neutrons, and it is found that 
the proton particle-hole excitation generating the 
$\gamma-$vibrational mode in $^{24}$Mg continues to play a key role in 
the near-drip-line nucleus $^{40}$Mg. 
The low-frequency octupole excitations are also investigated and 
the microscopic mechanism for the enhancement of transition strengths 
is discussed. 
} 
\maketitle
\section{Introduction}\label{intro}

Collective motion in unstable nuclei has raised a considerable 
interest both experimentally and theoretically. This is because
low-frequency modes of excitation are quite sensitive to the shell
structure near the Fermi level, and we can expect unique
excitation modes to emerge associated with the new spatial
structures such as neutron skins and the novel shell structures
that generate new regions of deformation~\cite{sto03}. 

In order to investigate
new kinds of excitation modes in exotic nuclei, 
the random-phase approximation (RPA) based on the self-consistent mean field
has been employed by many groups. 
(See Refs.~\cite{ben03,vre05,paa07} for
extensive lists of references concerning the self-consistent RPA and mean-field calculations. ) 
They are however largely restricted to spherical systems,
and the low-frequency excitation modes in deformed neutron-rich nuclei 
remain mostly unexplored. 

Recently, low-lying RPA modes in deformed
neutron-rich nuclei have been investigated by several
groups~\cite{yos05,nak05,ina06,sar98,mor06,urk01,hag04a,pen07,per08}.
These calculations, however, do not take into account the pairing
correlations, or rely on the BCS approximation for pairing (except for Ref.~\cite{per08}), 
which is inappropriate for
describing the pairing correlations in drip line nuclei due to the
unphysical nucleon gas problem~\cite{dob84}.
Quite recently, we have developed a new framework of the self-consistent 
deformed quasiparticle-RPA (QRPA) based on the
Skyrme-Hartree-Fock-Bogoliubov (HFB) mean field~\cite{yos08}. 

Presently, small excitation energies of the first $2^{+}$ state and 
striking enhancements of $B(E2;0_{1}^{+} \to 2_{1}^{+})$ 
in $^{32}$Mg~\cite{gui84,mot95} and $^{34}$Mg~\cite{yon01,chu05,ele06} 
are under lively discussions 
in connection with the onset of quadrupole deformation, 
the breaking of $N=20$ spherical magic number, 
the pairing correlation and the continuum coupling effects~\cite{pov87,war90,fuk92,uts99,yam04}. 
In order to get clear understanding of the nature of quadrupole deformation 
and pairing correlations, 
it is strongly desired to explore, both experimentally and theoretically,
excitation spectra of these nuclei toward a drip line~\cite{gad07,bau07,ter97,cau98,rei99,rod02,yos06}.

In this paper, we apply the new calculation scheme to the 
low-frequency excitation modes in neutron-rich Magnesium isotopes 
close to the drip line, 
and investigate the microscopic mechanism 
of the excitation modes uniquely appearing in deformed neutron-rich nuclei. 

The paper is organized as follows: 
In Sec.~\ref{method}, the deformed Skyrme-HFB + QRPA method is recapitulated. 
In Sec.~\ref{result}, results of numerical analysis of the low-lying excitation modes 
in deformed neutron-rich Magnesium isotopes are presented. 
Finally, summary is given in Sec.~\ref{summary}. 

\section{Method}\label{method}

\subsection{Skyrme-HFB in cylindrical coordinates}
In order to describe simultaneously the nuclear deformation 
and the pairing correlations including the unbound quasiparticle states, 
we solve the HFB equations~\cite{dob84,bul80}
\begin{align}
\begin{pmatrix}
h^{q}(\boldsymbol{r},\sigma)-\lambda^{q} & \tilde{h}^{q}(\boldsymbol{r},\sigma) \\
\tilde{h}^{q}(\boldsymbol{r},\sigma) & -(h^{q}(\boldsymbol{r},\sigma)-\lambda^{q})
\end{pmatrix}
\begin{pmatrix}
\varphi^{q}_{1,\alpha}(\boldsymbol{r},\sigma) \\
\varphi^{q}_{2,\alpha}(\boldsymbol{r},\sigma)
\end{pmatrix} \notag \\
= E_{\alpha}
\begin{pmatrix}
\varphi^{q}_{1,\alpha}(\boldsymbol{r},\sigma) \\
\varphi^{q}_{2,\alpha}(\boldsymbol{r},\sigma)
\end{pmatrix} \label{HFB_equation}
\end{align}
in coordinate space using cylindrical coordinates $\boldsymbol{r}=(\rho,z,\phi)$.
We assume axial and reflection symmetries.
Here, $q=\nu$ (neutron) or $\pi$ (proton).
For the mean-field Hamiltonian $h$, we employ the SkM* interaction~\cite{bar82}. 
Details for expressing the densities and currents in the cylindrical coordinate
representation can be found in Ref.~\cite{ter03}.
The pairing field is treated by using the density-dependent contact
interaction~\cite{cha76},
\begin{equation}
v_{pair}(\boldsymbol{r},\boldsymbol{r}^{\prime})=\dfrac{1-P_{\sigma}}{2}
\left[ t_{0}^{\prime}+\dfrac{t_{3}^{\prime}}{6}\varrho_{0}^{\gamma}(\boldsymbol{r}) \right]
\delta(\boldsymbol{r}-\boldsymbol{r}^{\prime}). \label{pair_int}
\end{equation}
where $\varrho_{0}(\boldsymbol{r})$ denotes the isoscalar density of the ground state 
and $P_{\sigma}$ the spin exchange operator.
Assuming time-reversal symmetry and reflection symmetry with respect to the $x-y$ plane,
we have to solve for positive $\Omega$ and positive $z$ only, 
$\Omega$ being the $z-$component of the angular momentum $j$. 
We use the lattice mesh size $\Delta\rho=\Delta z=0.6$ fm and a box
boundary condition at $\rho_{\mathrm{max}}=9.9$ fm, $z_{\mathrm{max}}=12$ fm. 
The differential operators are represented by use of the 11-point formula of Finite Difference Method. 
Because the parity and $\Omega$ are good quantum numbers in the present calculation scheme, 
we have only to diagonalize the HFB Hamiltonian (\ref{HFB_equation}) for each $\Omega^{\pi}$ sector. 
The quasiparticle energy is cut off at $E_{\mathrm{qp,cut}}=60$ MeV
and the quasiparticle states up to $\Omega^{\pi}=15/2^{\pm}$ are included.

The pairing strength parameter $t_{0}^{\prime}$ is
determined so as to reproduce the experimental pairing gap of
$^{34}$Mg ($\Delta_{\mathrm{exp}}=1.7$ MeV) obtained by the
three-point formula~\cite{sat98}.
The strength $t_{0}^{\prime}=-295$MeV$\cdot$fm$^{3}$ for the
mixed-type interaction ($t_{3}^{\prime}=-18.75t_{0}^{\prime}$)~\cite{ben05} 
with $\gamma=1$ leads to the pairing gap
$\langle \Delta_{\nu}\rangle=1.71$ MeV in $^{34}$Mg.

\subsection{Quasiparticle-basis QRPA}
Using the quasiparticle basis obtained
as the self-consistent solution of the HFB equations (\ref{HFB_equation}),
we solve the QRPA equation in the matrix formulation~\cite{row70}
\begin{equation}
\sum_{\gamma \delta}
\begin{pmatrix}
A_{\alpha \beta \gamma \delta} & B_{\alpha \beta \gamma \delta} \\
B_{\alpha \beta \gamma \delta} & A_{\alpha \beta \gamma \delta}
\end{pmatrix}
\begin{pmatrix}
f_{\gamma \delta}^{n} \\ g_{\gamma \delta}^{n}
\end{pmatrix}
=\hbar \omega_{n}
\begin{pmatrix}
1 & 0 \\ 0 & -1
\end{pmatrix}
\begin{pmatrix}
f_{\alpha \beta}^{n} \\ g_{\alpha \beta}^{n}
\end{pmatrix} \label{eq:AB1}.
\end{equation}
The residual interaction in the particle-hole (p-h) channel appearing
in the QRPA matrices $A$ and $B$ is
derived from the Skyrme density functional. 
We neglect the spin-orbit interaction term
$C_{t}^{\nabla J}$ as well as the
Coulomb interaction to reduce the computing time.
We also drop the so-called $``{J}^{2}"$ term $C_{t}^{T}$ both in 
the HFB and QRPA calculations. 
The residual interaction in the
particle-particle (p-p) channel is derived from the pairing
functional constructed with the density-dependent contact
interaction (\ref{pair_int}).

Because the full self-consistency between the static mean-field
calculation and the dynamical calculation is broken by the above
neglected terms, we renormalize the residual interaction in the
p-h channel by an overall factor $f_{\mathrm{ph}}$ to get the spurious
$K^{\pi}=0^{-}$ and $1^{-}$  modes (representing the
center-of-mass motion), and $K^{\pi}=1^{+}$ mode (representing the
rotational motion in deformed nuclei) at zero energy 
($v_{\mathrm{ph}} \rightarrow f_{\mathrm{ph}}\cdot v_{\mathrm{ph}}$). 
We cut the two-quasiparticle
(2qp) space at $E_{\alpha}+E_{\beta} \leq 60$ MeV due to the
excessively demanding computer memory size and computing time
for the model space consistent with that adopted in the HFB
calculation; $2 E_{\mathrm{qp,cut}}=120$ MeV.
Accordingly, we need another factor $f_{\mathrm{pp}}$ for the p-p channel. 
We determine this factor such that the spurious $K^{\pi}=0^{+}$ mode associated
with the particle number fluctuation (representing the pairing
rotational mode) appears at zero energy 
($v_{\mathrm{pp}} \rightarrow f_{\mathrm{pp}}\cdot v_{\mathrm{pp}}$). 
(See Ref.~\cite{yos08} for details of determination of the normalization factors.)

In the present calculation, 
the dimension of the QRPA matrix (\ref{eq:AB1}) 
for the quadrupole $K^{\pi}=0^{+}$ excitation in $^{40}$Mg 
is about 16 400, and the memory size is 24.2 GB. 
The normalization factors are $f_{\mathrm{ph}}=1.106$, and $f_{\mathrm{pp}}=1.219$. 

In terms of the nucleon annihilation and creation operators 
in the coordinate representation, 
$\hat{\psi}(\boldsymbol{r}\sigma)$ 
and $\hat{\psi}^{\dagger}(\boldsymbol{r}\sigma)$, 
the quadrupole operator is represented as 
\begin{equation}
\hat{Q}_{2K}
=\sum_{\sigma}\int d\boldsymbol{r}r^{2}Y_{2K}(\hat{r})
\hat{\psi}^{\dagger}(\boldsymbol{r}\sigma)
\hat{\psi}(\boldsymbol{r}\sigma).
\end{equation} 
The intrinsic matrix elements 
$\langle n|\hat{Q}_{2K}|0 \rangle$ 
of the quadrupole operator between the excited state $|n \rangle$ 
and the ground state $|0\rangle$ are given by
\begin{equation}
\langle n |\hat{Q}_{2K}|0 \rangle=\sum_{\alpha \beta}
Q_{2K,\alpha \beta}^{(\mathrm{uv})} 
(f_{\alpha \beta}^{n}+g_{\alpha \beta}^{n})
=\sum_{\alpha \beta}M_{2K,\alpha \beta}^{(\mathrm{uv})} 
\label{eq:matrix_element}.
\end{equation}
The explicit expression of $Q_{2K,\alpha \beta}^{(\mathrm{uv})}$ is 
given in Ref.~\cite{yos06}. 
The neutron (proton) matrix element $M_{\nu}$ ($M_{\pi}$) is defined 
\begin{equation}
M_{\nu}=\sum_{\alpha\beta \in \nu}M_{2K,\alpha\beta}^{(\mathrm{uv})}, \hspace{0.5cm}
M_{\pi}=\sum_{\alpha\beta \in \pi}M_{2K,\alpha\beta}^{(\mathrm{uv})}.
\end{equation}

\subsection{Elimination of the spurious center-of-mass modes}
It is known that the self-consistent RPA, 
if the same effective interaction 
or the same energy density functional is used exactly both for 
the ground state and for the excited state, 
restores translational invariance~\cite{RS}. 
Because the present calculation scheme is not fully self-consistent, 
the calculated states $|n \rangle$ for $K^{\pi}=0^{-}$ and $1^{-}$ excitations 
may contain the spurious component of the center-of-mass motion. 
In order to separate the intrinsic excitations from the spurious excitation, 
the ``physical" states $| \tilde{n} \rangle$ are assumed 
\begin{equation}
\hat{X}_{\tilde{n}}=\hat{X}_{n}-\chi_{P}^{n}\hat{P}-\chi_{R}^{n}\hat{R},
\end{equation}
where $\hat{X}_{n}$ is an annihilation operator of the calculated RPA mode, 
$\hat{R}$ and $\hat{P}$ are the coordinate and momentum operators of the whole nucleus 
($[\hat{R},\hat{P}]=i\hbar$). 
The coefficients $\chi$ are considered to be small 
because the spurious component is expected to reasonably decouple from 
the physical solutions. 
The physical states satisfy the following conditions. 
\begin{enumerate}
\item
The vacuum condition:
\begin{equation}
\hat{X}_{\tilde{n}}|0\rangle =0
\end{equation}
\item
The decoupling condition: 
\begin{equation}
\langle 0| [\hat{X}_{\tilde{n}},\hat{P}] |0\rangle =0, 
\langle 0| [\hat{X}_{\tilde{n}},\hat{R}] |0\rangle =0
\end{equation}
\end{enumerate}
These conditions determine the coefficients $\chi_{P}^{n}$, $\chi_{R}^{n}$;
\begin{equation}
\chi_{P}^{n}=i\langle 0|[\hat{X}_{n},\hat{R}]|0\rangle/\hbar, 
\chi_{R}^{n}=-i\langle 0|[\hat{X}_{n},\hat{P}]|0\rangle/\hbar.
\end{equation}
The orthonormality of the physical states $|\tilde{n}\rangle$ satisfies 
in first order of the correction coefficient $\chi$;
\begin{align}
\langle \tilde{n}|\tilde{m}\rangle =
& \langle 0|[\hat{X}_{\tilde{n}},\hat{X}^{\dagger}_{\tilde{m}}]|0\rangle \notag \\
=& \langle 0|[\hat{X}_{n},\hat{X}^{\dagger}_{m}]|0\rangle 
-i\hbar(\chi_{P}^{n}\chi_{R}^{m *}-\chi_{R}^{n}\chi_{P}^{m *}) \notag \\
\simeq & \delta_{n,m}.
\end{align}

In the actual calculations, the correction coefficient $\chi^{2}$ 
is at most of order $10^{-5}$.
Separation of the spurious modes is also proposed in Ref.~\cite{nak07} in a similar way 
to the transition density.

The explicit expressions 
for calculating the matrix elements of the octupole transition operator 
are given in Appendix.

\section{Results and discussion}\label{result}
\subsection{Ground state properties}

\begin{table}[b]
\begin{center}
\caption{Ground state properties of $^{34,36,38,40}$Mg obtained by the deformed HFB calculation 
with the SkM* interaction and the mixed-type pairing interaction. Chemical potentials, 
deformation parameters, average pairing gaps, root-mean-square radii for 
neutrons and protons are listed. The average pairing gaps of protons are zero in these isotopes. 
The average pairing gap is defined 
$\langle \Delta \rangle_{q}
=-\int d\boldsymbol{r}\tilde{h}\tilde{\varrho}/\int d\boldsymbol{r}\tilde{\varrho}$.}
\label{GS}
\begin{tabular}{ccccc}
\hline\noalign{\smallskip}
 & $^{34}$Mg & $^{36}$Mg & $^{38}$Mg & $^{40}$Mg  \\
\noalign{\smallskip}\hline\noalign{\smallskip}
$\lambda_{\nu}$ (MeV) & $-4.16$ & $-3.24$ & $-2.41$ & $-1.56$ \\
$\lambda_{\pi}$ (MeV) & $-19.8$ & $-21.0$ & $-23.7$ & $-24.4$ \\
$\beta_{2}^{\nu}$ & 0.35 & 0.31 & 0.29 & 0.28 \\
$\beta_{2}^{\pi}$ & 0.41 & 0.39 & 0.38 & 0.36 \\
$\langle \Delta \rangle_{\nu}$ (MeV) & 1.71 & 1.71 & 1.64 & 1.49 \\
$\sqrt{\langle r^{2} \rangle_{\nu}}$ (fm) & 3.51 & 3.59 & 3.67 & 3.76 \\
$\sqrt{\langle r^{2} \rangle_{\pi}}$ (fm) & 3.16 & 3.18 & 3.20 & 3.22 \\
\noalign{\smallskip}\hline
\end{tabular}
\end{center}
\end{table}

\begin{figure}[t]
\begin{center}
\resizebox{0.23\textwidth}{!}{
\includegraphics{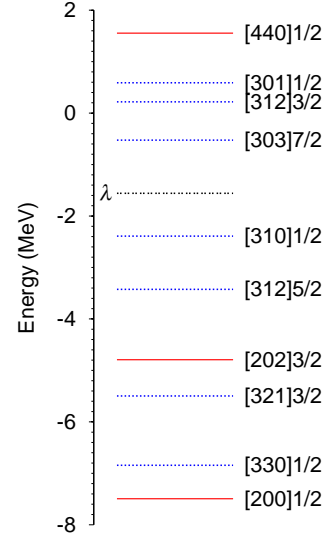}}
\caption{Neutron single-particle levels in $^{40}$Mg 
labeled with the asymptotic quantum numbers $[Nn_{3}\Lambda]\Omega$. 
The solid and dotted lines stand for the positive and negative parities. The chemical potential 
$\lambda$ is indicated by the two-dotted line.
}
\label{40Mg_spe}
\end{center}
\end{figure}

\begin{figure*}[t]\sidecaption
\resizebox{0.6\textwidth}{!}{
\includegraphics{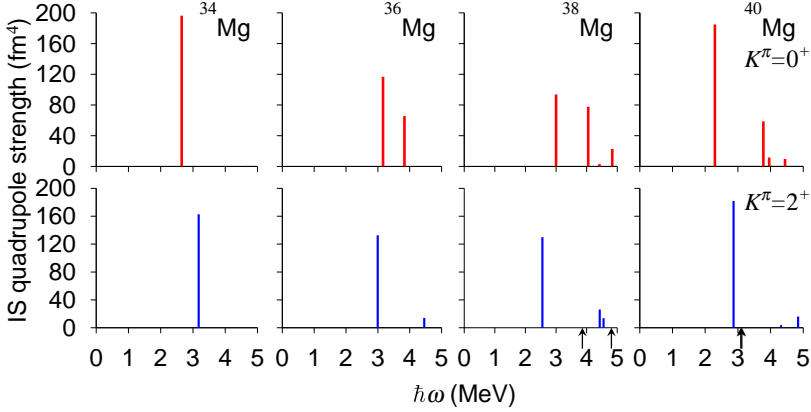}}
\caption{Intrinsic isoscalar quadrupole transition strengths in $^{34,36,38,40}$Mg 
for $K^{\pi}=0^{+}$ (upper panel) and $2^{+}$ (lower panel) excitations. 
The arrows indicate the neutron emission threshold energies. 
The one-neutron emission threshold energy is 
$E_{\mathrm{th},1n}=3.86$ MeV and 3.09 MeV, 
and the two-neutron emission threshold energy is 
$E_{\mathrm{th},2n}$= 4.81 MeV and 3.12 MeV in $^{38}$Mg and $^{40}$Mg, respectively. 
}
\label{Mg_strength}
\end{figure*}

In Table~\ref{GS}, the ground state properties are summarized. 
The neutron-rich Magnesium isotopes under investigation are prolately deformed. 
This is consistent with the results calculated 
using the Skyrme SIII interaction except for $^{34}$Mg~\cite{ter97}. 
The Gogny-HFB calculation suggested that $^{34}$Mg is prolately deformed 
but soft against $\beta$ deformation, and the shape coexistence in $^{38,40}$Mg~\cite{rod02}. 
We can see that the neutron skin develops as approaching the drip line; 
the difference in neutron and proton radii 
$\sqrt{\langle r^{2} \rangle_{\nu}}-\sqrt{\langle r^{2} \rangle_{\pi}}=0.35$ fm 
in $^{34}$Mg changes to 0.54 fm in $^{40}$Mg. 

In Fig.~\ref{40Mg_spe}, we show the neutron single-particle levels in $^{40}$Mg. 
The single-particle states are obtained by
rediagonalizing the self-consistent single-particle
Hamiltonian $h[\varrho,\tilde{\varrho}]$ of Eq.~(\ref{HFB_equation}). 
According to the present calculation employing the SkM* Skyrme density functional 
and the mixed-type pairing density functional, 
$^{40}$Mg is located close to the neutron drip line.

\subsection{Quadrupole vibrations}
Figure~\ref{Mg_strength} shows the intrinsic 
isoscalar quadrupole transition strengths in neutron-rich Mg isotopes 
for $K^{\pi}=0^{+}$ and $2^{+}$ excitations. 
For the $K^{\pi}=0^{+}$ excitation, 
we can see a prominent peak 
possessing about 30 and 23 in Weisskopf unit below the threshold energy 
in $^{34}$Mg and $^{40}$Mg. 
If we assume the strong deformation limit~\cite{BM2}, 
these intrinsic isoscalar transition strengths correspond to 
the transition strengths from the ground $0^{+}_{1}$ state 
to the $2^{+}_{\beta}$ state built on the excited $K^{\pi}=0^{+}$ state, 
and those from the excited $K^{\pi}=0^{+}$ state to the $2_{1}^{+}$ state 
built on the ground $0^{+}_{1}$ state in the laboratory frame. 

On the other hand, we obtain the collective state in all of the isotopes 
for the $K^{\pi}=2^{+}$ excitation. 

\begin{table}[t]
\begin{center}
\caption{Ratios of the neutron and proton matrix elements $M_{\nu}/M_{\pi}$ 
divided by $N/Z$ for the lowest excited states 
in Mg isotopes for the quadrupole $K^{\pi}=0^{+}$ and $2^{+}$ excitations.}
\label{ratio}
\begin{tabular}{ccccc}
\hline\noalign{\smallskip}
 & $^{34}$Mg & $^{36}$Mg & $^{38}$Mg & $^{40}$Mg  \\
\noalign{\smallskip}\hline\noalign{\smallskip}
$K^{\pi}=0^{+}$ & 1.57 & 1.58 & 1.82 & 1.91 \\
$K^{\pi}=2^{+}$ & 1.41 & 1.41 & 1.55 & 1.79 \\
\noalign{\smallskip}\hline
\end{tabular}
\end{center}
\end{table}

In Table~\ref{ratio}, we summarize the ratio of the matrix elements for neutrons and protons 
normalized by that of the neutron and proton numbers, $(M_{\nu}/M_{\pi})/(N/Z)$, 
for the lowest excitation modes. 
As approaching the neutron drip line, 
the contribution of the neutron excitation becomes large. 
This is one of the unique features of the excitation modes in drip-line nuclei 
and it is understood as follows: 
In drip-line nuclei, the neutron 2qp excitations dominantly take place outside of the nuclear surface. 
Therefore, the transition strengths of the 2qp excitation of neutrons become large. 
The proton p-h excitations, however, concentrate in the surface region. 
Consequently, coupling of the excitations between neutrons and protons becomes smaller, 
and the transition strengths of neutrons ($M_{\nu}^{2}$) 
and protons ($M_{\pi}^{2}$) become extremely asymmetric. 
In Sec.~\ref{0+mode}, 
we discuss in detail the microscopic structure of the low-frequency $K^{\pi}=0^{+}$ modes, 
and show the spatial structure of the excitations of neutrons and protons.

\subsubsection{Soft $K^{\pi}=0^{+}$ modes}\label{0+mode}
\begin{figure}[t]
\begin{center}
\resizebox{0.41\textwidth}{!}{
\includegraphics{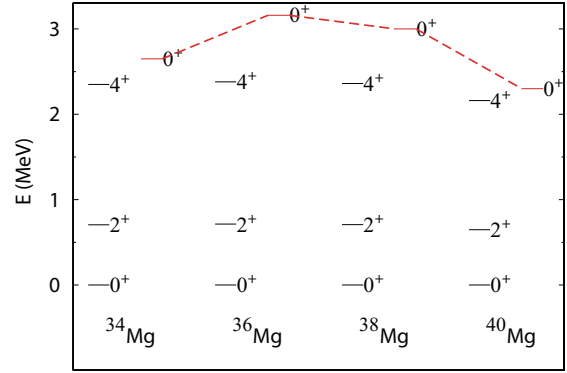}}
\caption{Low-excitation energy spectra of $^{34,36,38,40}$Mg.
}
\label{Mg_level}
\end{center}
\end{figure}

In Fig.~\ref{Mg_level}, we show the low-lying excitation spectra for the $K^{\pi}=0^{+}$ states. 
Here excitation energies are evaluated by~\cite{EG}
\begin{equation}
E(I,K)=\hbar \omega_{\mathrm{RPA}}+\frac{\hbar^{2}}{2\mathcal{J}_{\mathrm{TV}}}(I(I+1)-K^{2}),
\end{equation}
in terms of the vibrational frequencies $\omega_{\mathrm{RPA}}$ 
and the Thouless-Valatin moment of inertia 
$\mathcal{J}_{\mathrm{TV}}$ 
calculated microscopically by the QRPA as described in Ref.~\cite{yos08b}.
As we can see in this figure, 
appearance of the soft $K^{\pi}=0^{+}$ modes is quite sensitive to the neutron number. 

In Ref.~\cite{yos08b}, we have discussed
the generic feature of the low-lying $K^{\pi}=0^{+}$ modes in deformed
neutron-rich nuclei:
In a deformed system where the up-sloping
oblate-type and the down-sloping prolate-type orbitals exist near
the Fermi level, one obtains a low-lying mode possessing enhanced
strengths both for the quadrupole p-h transition and for the
quadrupole p-p (pair) transition induced by the pairing
fluctuations. 
The up-sloping and down-sloping orbitals have quadrupole moments with opposite signs. 

Generation mechanism of the soft $K^{\pi}=0^{+}$ mode in deformed neutron-rich nuclei is 
understood essentially by the schematic two-level model in Ref.~\cite{BM2}.
They consider the case where only two $\lambda \bar{\lambda}$ components 
are present in the wave functions both of the ground $0^{+}_{1}$ and 
of the excited $0^{+}_{2}$ states;
\begin{subequations}
\begin{align}
|K^{\pi}=0^{+}_{1}\rangle &= \dfrac{a}{\sqrt{a^2+b^2}}|\lambda_{1}\bar{\lambda}_{1}\rangle + 
\dfrac{b}{\sqrt{a^2+b^2}}|\lambda_{2}\bar{\lambda}_{2}\rangle \\
|K^{\pi}=0^{+}_{2}\rangle &= -\dfrac{b}{\sqrt{a^2+b^2}}|\lambda_{1}\bar{\lambda}_{1}\rangle + 
\dfrac{a}{\sqrt{a^2+b^2}}|\lambda_{2}\bar{\lambda}_{2}\rangle.
\end{align}
\end{subequations}
The transition matrix element for the quadrupole operator is then 
\begin{equation}
\langle 0^{+}_{1}|\hat{Q}_{20}|0^{+}_{2}\rangle =
\dfrac{2ab}{a^{2}+b^{2}}[ \langle \lambda_{1}|\hat{Q}_{20}|\lambda_{1}\rangle 
- \langle \lambda_{2}|\hat{Q}_{20}|\lambda_{2}\rangle ]
\end{equation}
and it is proportional to the difference in the 
quadrupole moments of the individual orbitals composing the $0^{+}$ states.
In the case that the quadrupole moments of the orbitals have opposite signs to each other, 
this matrix element becomes large. 
This situation is realized in the level crossing region, where the up- and down-sloping 
orbitals exist. 
As the number of components increases in the QRPA calculations, 
the wave function becomes more complicated. 
It is discussed in Ref.~\cite{yos08b}. 

\begin{figure}[t]
\begin{center}
\resizebox{0.41\textwidth}{!}{
\includegraphics{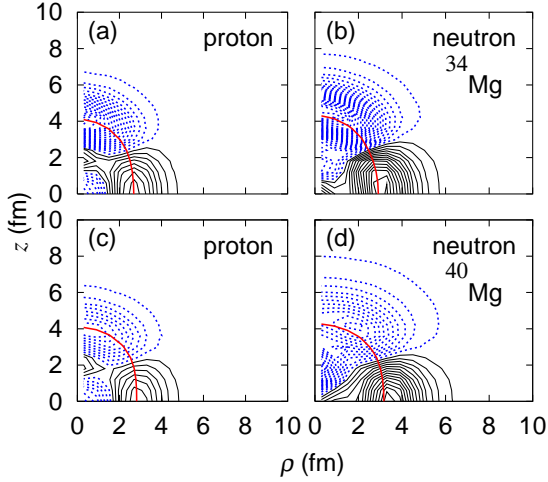}
}
\caption{Transition densities of protons (left) and neutrons (right) to 
the $K^{\pi}=0^{+}$ states at 2.65 MeV in $^{34}$Mg (upper) and 
at 2.30 MeV in $^{40}$Mg (lower). 
Solid and dotted lines indicate positive 
and negative transition densities, and 
the contour lines are plotted at intervals of $3 \times 10^{-4}$ fm$^{-3}$. 
The thick solid lines indicate the neutron and proton half densities of the ground state. 
They are 0.054 and 0.036 fm$^{-3}$ for neutrons and protons, respectively in $^{34}$Mg, 
and 0.055 and 0.032 fm$^{-3}$ in $^{40}$Mg.
}
\label{40Mg_trans_density}
\end{center}
\end{figure}

In what follows, we investigate the microscopic structure of 
the low-lying $K^{\pi}=0^{+}$ states, 
and discuss sensitivity to the location of the Fermi level of neutrons. 
Because the deformation properties of the Mg isotopes under investigation 
are not very different, 
Figure~\ref{40Mg_spe} is used for understanding the shell structure around the 
Fermi level. 

In $^{34}$Mg, we obtain the collective $K^{\pi}=0^{+}$ mode at 2.65MeV~\cite{yos08}.  
The transition strength is enhanced by 10.6 times as compared to the
unperturbed transition strength.  
This mode is generated by many 2qp excitations, 
and among them the 2qp configurations of 
$(\nu[202]3/2)^{2}$ and $(\nu[321]3/2)^{2}$ have main
contributions with weights of 0.44 and 0.34, respectively. 
The chemical potential is located between these two levels. 
They are a up-sloping and a down-sloping orbitals, respectively. 

In $^{36}$Mg, we obtain two weak-collective states. 
The second $K^{\pi}=0^{+}$ state at 3.84 MeV is mainly generated
by the 2qp excitations of $(\nu[321]3/2)^{2}$ and
$(\nu[312]5/2)^{2}$ with weights of 0.48 and 0.35. 
The lowest $K^{\pi}=0^{+}$ state at 3.17 MeV is analogous to the collective
state in $^{34}$Mg: This is mainly generated by the 2qp
excitations of $(\nu[202]3/2)^{2}$ and $(\nu[310]1/2)^{2}$ with 
weights of 0.58 and 0.21, which are a up-sloping and a 
down-sloping levels, respectively. 

We also obtain two weak-collective states in $^{38}$Mg at 3.00 MeV and 4.05 MeV. 
The lower state has a similar structure to the lowest state in $^{36}$Mg: 
This is mainly generated by the 2qp excitations of $(\nu[202]3/2)^{2}$ and 
$(\nu[310]1/2)^{2}$ with weights of 0.11 and 0.59. 
Furthermore, the 2qp excitation of $(\nu[312]5/2)^{2}$ 
has an appreciable contribution of 0.20. 
This excitation, however, acts destructively to the above excitations. 
Therefore the transition strength to the lowest state is not enhanced. 
The second $K^{\pi}=0^{+}$ state is located just above the continuum threshold. 
This state is mainly generated by the excitations of 
$(\nu[303]7/2)^{2}$ and $(\nu[312]5/2)^{2}$ with weights of 0.38 and 0.34. 

In $^{40}$Mg, we can see a prominent peak at 2.30 MeV. 
This state is mainly generated by the excitations of $(\nu[310]1/2)^{2}$ and $(\nu[303]7/2)^{2}$ 
with weights of 0.50 and 0.29. 
The $\nu[303]7/2$ orbital is a up-sloping level stemming from the $1f_{7/2}$ orbital. 
Furthermore, many 2qp excitations of neutrons coherently participate 
in generating this soft $K^{\pi}=0^{+}$ mode.

The transition densities to the 
soft $K^{\pi}=0^{+}$ modes in $^{34}$Mg and $^{40}$Mg are shown 
in Fig.~\ref{40Mg_trans_density}. 
Inside and around the nuclear surface denoted by the thick lines, 
neutrons and protons coherently oscillate 
along the symmetry axis ($z-$axis), indicating the $\beta$ vibration. 
Furthermore, in neutrons only, we can see a spatially extended structure. 
The enhanced transition strength of neutrons is due to 
this spatial extension of the quasiparticle wave functions. 

\begin{figure}[t]
\begin{center}
\resizebox{0.22\textwidth}{!}{
\includegraphics{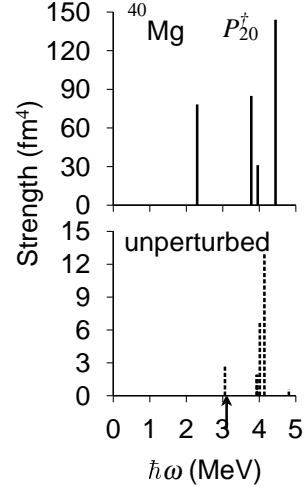}}
\caption{Transition strengths for the quadrupole-pair excitation in $^{40}$Mg. 
The unperturbed 2qp transition strengths are shown in the lower panel.}
\label{40Mg_pair_strength}
\end{center}
\end{figure}

In order to see another interesting feature of the soft $K^{\pi}=0^{+}$ mode 
in $^{40}$Mg, we show in Fig.~\ref{40Mg_pair_strength} 
the strength distributions of the quadrupole-pair transition 
defined by the operator
\begin{equation}
\hat{P}^{\dagger}_{20}=\int d\boldsymbol{r} r^{2}Y_{20}(\hat{r})
\hat{\psi}^{\dagger}_{\nu}(\boldsymbol{r}\uparrow)\hat{\psi}^{\dagger}_{\nu}(\boldsymbol{r}\downarrow).
\end{equation}

The transition strength to the lowest state is enhanced 
about 28.4 times of the transition strength of the 2qp excitation $(\nu[310]1/2)^{2}$. 
The lowest $K^{\pi}=0^{+}$ state in $^{40}$Mg thus has a collective nature both in the 
p-h and p-p (pairing) channels. 

\subsubsection{$K^{\pi}=2^{+}$ modes}\label{2+mode}
Next, we investigate the microscopic structure of the $K^{\pi}=2^{+}$ modes
appearing at around 3 MeV. 

In $^{34}$Mg, 
two quasiparticle levels near the Fermi level, [202]3/2 and [321]3/2, 
play a dominant role in generating the $K^{\pi}=2^{+}$ mode 
at 3.18 MeV: 
2qp excitations of $\nu[202]3/2 \otimes \nu[220]1/2$ 
and $\nu[310]1/2 \otimes \nu[321]3/2$ have large contribution 
with weights of 0.51 and 0.14. 

Because a $[312]5/2$ level is located close to the Fermi level 
in $^{36}$Mg, 
the 2qp excitation of $\nu[310]1/2 \otimes \nu[312]5/2$ 
becomes a dominant component with a weight of 0.54. 
This component is not large in $^{34}$Mg, 
which has 5\% of the contribution. 
In $^{38}$Mg, both of the $[312]5/2$ and $[310]1/2$ levels are 
close to the Fermi level. 
Therefore, the 2qp excitation of $\nu[310]1/2 \otimes \nu[312]5/2$ 
has larger contribution possessing a weight of 0.75. 
The 2qp excitation of $\nu[310]1/2 \otimes \nu[321]3/2$ still 
have an appreciable contribution with weights of 0.14 and 0.06
in $^{36}$Mg and $^{38}$Mg.

In $^{40}$Mg, 
the 2qp excitation of $\nu[310]1/2 \otimes \nu[312]5/2$ 
has a large contribution with a weight of 0.24 
similarly to the $K^{\pi}=2^{+}$ states in $^{36}$Mg and $^{38}$Mg. 
Furthermore, because the [303]7/2 level is close to the Fermi level, 
the 2qp excitation of $\nu[312]3/2 \otimes \nu[303]7/2$ 
becomes a dominant component possessing 37\% contribution. 

In all of the $K^{\pi}=2^{+}$ states in Mg isotopes 
under investigation, 
the proton p-h excitations also play an important role. 
Especially, the $\pi[211]3/2 \to \pi[211]1/2$ excitation 
has an appreciable contribution with weights of 
0.21, 0.16, 0.09 and 0.09 in $^{34,36,38,40}$Mg, respectively. 
It is noted that  
the $\gamma$-vibrational mode in $^{24}$Mg is mainly generated by
the neutron and proton p-h excitations of $[211]3/2\to[211]1/2$~\cite{yos08}.

\subsection{Octupole excitations}\label{negative-parity}

\begin{figure}[t]
\begin{center}
\resizebox{0.48\textwidth}{!}{
\includegraphics{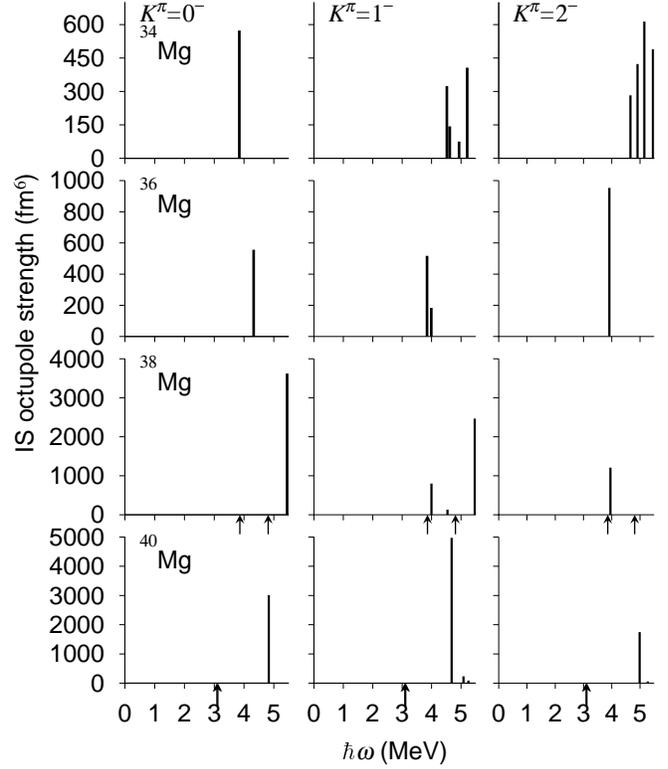}}
\caption{Intrinsic isoscalar octupole transition strengths in $^{34,36,38,40}$Mg 
for $K^{\pi}=0^{-}$ (left panel), $K^{\pi}=1^{-}$ (middle panel) 
and $2^{-}$ (right panel) excitations. 
}
\label{Mg_strength_oct}
\end{center}
\end{figure}

Figure~\ref{Mg_strength_oct} 
shows the intrinsic isoscalar octupole transition strengths. 
For the $K^{\pi}=3^{-}$ excitations, 
there are no peaks representing strengths greater 
than 1 W.u. (1 W.u. is about 70--90 fm$^{6}$ in $^{34-40}$Mg) 
in the low-excitation energy region below 5 MeV. 

Because the number of 2qp negative-parity excitations 
in this mass region is small 
as we can see the single-particle energy levels in Fig.~\ref{40Mg_spe}, 
none of the excited states shown in Fig.~\ref{Mg_strength_oct} are collective. 
For instance, 
both of the lowest $K^{\pi}=0^{-}$ states in $^{34}$Mg and $^{36}$Mg are 
generated by the $\nu[202]3/2 \otimes \nu[321]3/2$ excitation 
dominantly with a weight of 0.98. 
The unperturbed transition strength of this neutron 2qp excitation 
is about 20 fm$^{6}$. 
About 50\% of the transition matrix element come from 
the coupling to the giant resonances around 10 MeV and 30 MeV. 

In $^{38}$Mg and $^{40}$Mg, the lowest $K^{\pi}=0^{-}$ state is 
generated by the $\nu[440]1/2 \otimes \nu[310]1/2$ excitation 
with weights of 0.72 and 0.88, respectively. 
In $^{38}$Mg, the $\nu[202]3/2 \otimes \nu[321]3/2$ excitation 
still has a small contribution of 0.08. 
The transition strengths for the $K^{\pi}=0^{-}$ excitation 
in $^{38}$Mg and $^{40}$Mg are about 5 times 
larger than in $^{34,36}$Mg.
This is because the unperturbed transition strengths of the 
neutron 2qp excitations below 10 MeV become extremely large 
due to the spatial extension of the quasiparticle wave functions 
around the Fermi level; 
the transition strength of 
the $\nu[440]1/2 \otimes \nu[310]1/2$ excitation is 
313 fm$^{6}$ and 720 fm$^{6}$ in $^{38,40}$Mg. 
Since the $\nu[310]1/2$ orbital is a particle-like level 
in $^{38}$Mg, the p-h transition strength is smaller than in $^{40}$Mg. 
About 80\% of the transition matrix element come from the neutron 2qp 
excitations below 10 MeV, 
and coupling to the giant resonance is small. 
In drip-line nuclei, the proton contribution is also small 
as in the case for the quadrupole excitations. 
In $^{40}$Mg, the proton transition strength 
$B(E3; 0^{+}_{\mathrm{gs}}\to 3^{-}_{K^{\pi}=0^{-}})$ 
has only 5.0 $e^{2}$fm$^{6}$ 
whereas it has 33 $e^{2}$fm$^{6}$ in $^{34}$Mg. 

\section{Summary}\label{summary}
We have investigated the microscopic structure of the 
low-lying excitation modes systematically 
in neutron-rich Magnesium isotopes close to the drip line 
by means of the deformed Skyrme-QRPA method. 

In the spherical neutron-rich nuclei, the effect of the dynamical pairing 
has been investigated in detail in Refs.~\cite{mat01,mat05,paa03}. 
In Ref.~\cite{yos08b}, we showed 
the importance of the dynamical pairing in neutron-rich deformed systems. 
In a deformed system where the up- and down-sloping orbitals exist near the Fermi level, 
one obtains the low-lying mode possessing extremely enhanced strengths both 
for the quadrupole p-h transition and for the quadrupole p-p (pair) 
transition induced by the pairing fluctuation. 

In this article, we demonstrated the importance of the dynamical pairing 
in deformed drip-line system. 
Not only in $^{34}$Mg but in $^{40}$Mg, 
the soft $K^{\pi}=0^{+}$ modes are generated by the 
neutron 2qp excitations of 
the up-sloping and down-sloping orbitals. 
And we found that 
the transition strength to the soft $K^{\pi}=0^{+}$ modes 
is much enhanced not only for the quadrupole p-h excitation 
but also for the quadrupole p-p excitation. 

We obtained the collective $K^{\pi}=2^{+}$ excitations 
in all of the isotopes under investigation. 
As the neutron number increases, the contribution of 
neutron 2qp excitations to the $K^{\pi}=2^{+}$ mode changes gradually 
according to the location of the Fermi level of neutrons. 
Furthermore, 
we found that the proton p-h excitation of [211]3/2 $\to$ [211]1/2 
creating the $\gamma-$vibrational mode in $^{24}$Mg 
keeps playing an important role in generating 
the collective $K^{\pi}=2^{+}$ modes even in the near-drip-line nuclei. 

For the octupole excitations, 
we could not find any collective modes in the low-energy region. 
Despite the weak collectivity, 
the transition strengths are enhanced due to the microscopic mechanism: 
In $^{34}$Mg and $^{36}$Mg, coupling to the giant resonances at 
$1\hbar \omega_{0}$ and $3\hbar \omega_{0}$ 
gives rise to the enhancement of the transition strengths. 
In $^{38}$Mg and $^{40}$Mg close to the drip line, 
because the quasiparticle wave functions near the Fermi level 
have the spatially extended structure, the transition strengths 
of the neutron 2qp excitations become extremely large. 
However, the proton contribution becomes small 
in near-drip-line nuclei 
because the spatial overlap between 
the proton p-h excitations and neutron 2qp excitations 
is small. 

\begin{acknowledgement}
Stimulating discussions with K.~Matsuyanagi, T.~Nakatsukasa and K.~Yabana 
are greatly appreciated. 
The author is supported by the Special Postdoctoral Researcher Program of RIKEN.
The numerical calculations were performed on the NEC SX-8 supercomputer 
at the Yukawa Institute for Theoretical Physics, Kyoto University and 
the NEC SX-8R supercomputer 
at the Research Center for Nuclear Physics, Osaka University.
\end{acknowledgement}

\section*{Appendix}
\setcounter{equation}{0}
\renewcommand{\theequation}{A-\arabic{equation}}

The transition strengths between the RPA ground $|0 \rangle$ and 
the corrected ``physical" state $|\tilde{n}\rangle$ 
for the operator $\hat{O}$ are calculated as
\begin{equation}
|\langle \tilde{n}|\hat{O}|0\rangle |^{2}
=|\langle 0|[X_{\tilde{n}},\hat{O}]|0\rangle|^{2}.
\end{equation}
The transition matrix elements for the isoscalar octupole operator 
\begin{equation}
\hat{O}_{3K}=
\sum_{\sigma}\int d\boldsymbol{r} r^{3}Y_{3K}(\hat{r})
\hat{\psi}^{\dagger}(\boldsymbol{r}\sigma)
\hat{\psi}(\boldsymbol{r}\sigma)
\end{equation}
are given as 
\begin{align}
&\langle 0|[X_{\tilde{n}},\hat{O}_{30} ]|0\rangle \notag \\
&= \langle 0|[X_{n},\hat{O}_{30} ]|0\rangle \notag \\
& + 2\pi i\hbar \chi_{P_{z}}^{n} \sqrt{\dfrac{7}{16\pi}}
\int \rho d\rho dz \varrho_0(\rho,z)(6z^{2}-3\rho^{2})
\end{align}
for the $K^{\pi}=0^{-}$ states and 
\begin{align}
&\langle 0|[X_{\tilde{n}},\hat{O}_{31} ]|0\rangle \notag \\
& = \langle 0|[X_{n},\hat{O}_{31} ]|0\rangle \notag \\
& - 2\pi i\hbar \chi_{P_{\rho}}^{n} \sqrt{\dfrac{21}{64\pi}}
\int \rho d\rho dz \varrho_0(\rho,z)(4z^{2}-3\rho^{2}) 
\end{align}
for the $K^{\pi}=1^{-}$ states. Here  
\begin{align}
\chi_{P_{z}}^{n}
&=\dfrac{i}{\hbar} \dfrac{1}{A}\sum_{i=1}^{A}
\langle 0|[X_{n},\hat{z}_{i}]|0\rangle, \\
\chi_{P_{\rho}}^{n} 
&=\dfrac{i}{\hbar} \dfrac{1}{A}\sum_{i=1}^{A}
\langle 0|[X_{n},\hat{\rho}_{i}e^{i\phi}]|0\rangle.
\end{align}


\end{document}